\documentclass[twocolumn,showpacs,preprintnumbers,amsmath,amssymb,pra,aps,superscriptaddress]{revtex4-1}

\usepackage{graphicx}
\usepackage{multirow}
\usepackage{amsmath}
\usepackage{soul}
\usepackage{verbatim}
\usepackage{bm}
\usepackage{outlines}
\usepackage{siunitx}
\usepackage[space]{grffile}
\usepackage[colorlinks=true, linkcolor=blue, citecolor=gray]{hyperref}
\usepackage[capitalize]{cleveref}
\usepackage{xr}
\usepackage{color}
\usepackage{tikz}
\usetikzlibrary{calc, positioning}

\newcommand{\<}{\langle}
\renewcommand{\>}{\rangle}
\newcommand{\chat}{\hat{c}}
\newcommand{\chatdag}{\hat{c}^{\dag}}
\newcommand{\CC}{\mathbb{C}}
\newcommand{\DD}{\mathbb{D}}
\newcommand{\II}{\mathbb{I}}
\newcommand{\PP}{\mathbb{P}}
\newcommand{\RR}{\mathbb{R}}

\newcommand{\Ss}{\mathcal{S}}
\newcommand{\Trace}{\mathrm{Trace}}

\begin{document}
\graphicspath{{figures_resub/}}

\title{Low-cost error mitigation by symmetry verification}
\author{X. Bonet-Monroig}
\affiliation{Instituut-Lorentz, Universiteit Leiden, P.O. Box 9506, 2300 RA Leiden, The Netherlands}
\author{R. Sagastizabal}
\affiliation{QuTech, Delft University of Technology, P.O. Box 5046, 2600 GA Delft, The Netherlands}
\affiliation{Kavli Institute of Nanoscience, Delft University of Technology, P.O. Box 5046, 2600 GA Delft, The Netherlands}
\author{M. Singh}
\affiliation{QuTech, Delft University of Technology, P.O. Box 5046, 2600 GA Delft, The Netherlands}
\affiliation{Kavli Institute of Nanoscience, Delft University of Technology, P.O. Box 5046, 2600 GA Delft, The Netherlands}
\author{T.E. O'Brien}
\affiliation{Instituut-Lorentz, Universiteit Leiden, P.O. Box 9506, 2300 RA Leiden, The Netherlands}
\date{\today}
\begin{abstract}
We investigate the performance of error mitigation via measurement of conserved symmetries on near-term devices.
We present two protocols to measure conserved symmetries during the bulk of an experiment, and develop a third, zero-cost, post-processing protocol which is equivalent to a variant of the quantum subspace expansion.
We develop methods for inserting global and local symmetries into quantum algorithms, and for adjusting natural symmetries of the problem to boost the mitigation of errors produced by different noise channels.
We demonstrate these techniques on two- and four-qubit simulations of the hydrogen molecule (using a classical density-matrix simulator), finding up to an order of magnitude reduction of the error in obtaining the ground-state dissociation curve.
\end{abstract}
\maketitle

\section{Introduction}
Noisy, intermediate scale quantum (NISQ) devices have begun to appear in laboratories around the world.
These devices have performance rates around or just below the quantum error correction threshold~\cite{Ris15,Bar14,Deb16,Mon16,Ofe16}, but are lacking the number of qubits required for full fault-tolerant quantum computing.
This raises the open question of whether the upcoming generation of quantum computers will provide a quantum advantage over classical computers, and in which fields this might be achieved~\cite{Pre18,Mol18,Nei18}.
In particular, for the area of digital quantum simulation, it has been suggested that variational quantum eigensolvers~\cite{Per14} may be sufficiently low-cost to be performed on $\sim 50$ qubits~\cite{Bab18,Pou18,Ber18,Kiv18}.
Around this point, solving the many-body problem exactly becomes too challenging for classical computers, and a slight quantum edge might be available above current approximations.

In lieu of full error correction techniques, much attention is being turned to error mitigation techniques, which, although unscalable, promise modest improvements at low cost.
Previous work has focused on active error minimization, whereby data is obtained at artificially increased error rates and then extrapolated to zero~\cite{Tem17,Li17,Kan18,Ott18}, and on probabilistic error cancellation, where an ensemble of noisy circuits is applied such that they average to the target error-free circuit~\cite{Tem17,Endo17}.
More specific techniques have been developed for quantum simulation, and in particular for variational quantum eigensolvers.
A technique developed for exploring the low-energy excited subspace of a quantum system, the quantum subspace expansion, has been shown to have error mitigation as a side-effect~\cite{Mcc17,Col18}.

In this work we investigate error mitigation via verification of symmetries found in quantum circuits, in particular those in physical systems.
This is a low-cost version of the stabilizer parity checks ubiquitous in quantum error correction~\cite{Got09,Ter13}.
We develop multiple protocols to perform symmetry verification, both repeatedly throughout a quantum circuit and as a single post-processing step.
The latter can be related to a variant of the quantum subspace expansion~\cite{Mcc17}.
We study the sensitivity of symmetry verification to different noise channels, and demonstrate how it can be optimized by adding new symmetries and rotating existing symmetries to be more sensitive to local noise.

\section{Definitions}
In this section we cover some basic definitions to be used throughout this paper, and some details of quantum computing that may be skipped by the experienced reader.
We use the Pauli group on $N$ qubits $\PP^N=\{\II,X,Y,Z\}^{\otimes N}$ throughout this work.
These operators form a basis for operators $\hat{O}\in\CC^{2^N}$,
\begin{equation}
\hat{O}=\sum_{\hat{P}\in\PP^N}O_{\hat{P}}\hat{P},\label{eq:Pauli_decomp}
\end{equation}
with coefficients $O_{\hat{P}}$ in $\CC$.
If we choose $O_{\hat{P}}\in\RR$, this is then a basis for the $2^N\times 2^N$ Hermitian matrices.
We call such a basis decomposition of an operator $\hat{O}$ a Pauli decomposition.
Such a basis is orthogonal in the Frobenius norm:
\begin{equation}
||\hat{O}||_{F}=\sqrt{\Trace[\hat{O}^{\dag}\hat{O}]}.
\end{equation}
Elements $\hat{P}\in\PP^N$ have two eigenvalues $p=\pm 1$, with corresponding eigenspaces of dimension $2^{N-1}$, and projectors $\hat{M}_p=\frac{1}{2}(1 + pP)$ onto said eigenspaces.
The Pauli group has an additional property; if $\hat{P}\neq\II^{\otimes N}\in\PP^N$, $\hat{P}$ commutes with half of the elements of $\PP^N$ ($[\hat{P},\hat{Q}]=0$), and anticommutes with the other elements ($\{\hat{P},\hat{R}\}=0$).
This property can be extended to a general operator $\hat{O}$ - $[\hat{P},\hat{O}]=0$ ($\{\hat{P},\hat{O}\}=0$) if and only if $\hat{P}$ commutes (anticommutes) with each element of the Pauli decomposition of $\hat{O}$ (Eq.~\ref{eq:Pauli_decomp}).

A quantum computation consists of multiple experiments, each of which can be split into preparations, transformations, and measurements.
In the ideal case, a preparation generates a quantum state on a register of $N$ qubits, which is represented as a vector $|\phi\>$ in the complex vector space $\CC^{2^N}$.
Transformation consists of evolving this state to a new state $|\psi\>\in\CC^{2^N}$, which may be represented via a unitary operator $|\psi\>=U|\phi\>$ (with $U\in U(\CC^{2^N}), i.e. UU^{\dag}=U^{\dag}U=\II$).
Measurement consists of observing the quantum state $|\psi\>$ along some degree of freedom.
The degree of freedom is represented by a projector-valued measurement $\{\hat{M}_i\}$ for each possible observed value $i$, where $\sum_i\hat{M}_i=\II$, $\hat{M}_i^2=\hat{M}_i^\dag=\hat{M}_i$.
The observation records one such value $i$ at random with probability $p_i$ following the Born rule,
\begin{equation}
p_i=|\<\psi|\hat{M}|\psi\>|^2,
\end{equation}
and the state of the system collapses into $\hat{M}_i|\psi\>/p_i$.

In the presence of noise, the state of a qubit is instead given by a density matrix $\rho\in\DD^N$, where $\DD^N$ is the set of $2^N\times 2^N$ postive, trace $1$ matrices.
These are a generalized form of pure quantum states $|\psi\>$, which allow for statistical ensembles of pure states (the well-known adage being that preparing $\frac{1}{\sqrt{2}}(|0\>+|1\>)$ is strictly not the same as preparing $|0\>$ or $|1\>$ with $50\%$ probability).
For every pure state $|\psi\>$, the corresponding density matrix is the outer product $|\psi\>\<\psi|$, and the expectation value of an operator $\hat{O}$ may be calculated as
\begin{equation}
\<\psi|\hat{O}|\psi\>=\Trace[\hat{O}|\psi\>\<\psi|].
\end{equation}
We will use the latter notation throughout this paper, to be consistent with calculating expectation values on mixed states $\rho$ (where the standard bra-ket expectation value is no longer possible).
We will distinguish between operators $\hat{O}$ and density matrices $\rho$ by the use of hats.
Note that the trace of products of density matrices is also well defined, and has an obvious interpretation as the overlap between the density matrices, as for pure states
\begin{equation}
\Trace[|\psi\>\<\psi||\phi\>\<\phi|]=\<\psi|\phi\>\<\phi|\psi\>=|\<\phi|\psi\>|^2.
\end{equation}
Transformations and measurements of density matrices behave differently to those of pure states~\cite{Nie00}, but we will not need details of this in this work.

A quantum algorithm incurs a cost based on the number of qubits and coherence time required for quantum hardware to execute it.
This cost is usually increased by error mitigation protocols that require additional gates or ancilla qubits.
However, these are in general low cost compared to the overhead required for full quantum error correction.
Indeed, some error mitigation protocols require no additional quantum hardware or circuitry, hence we define them as `zero-cost'.
Such protocols may require repetition of the algorithm in order to estimate expectation values $\Trace[\hat{O}\rho]$, but this may be offset by parallelizing across multiple quantum devices.
This cost metric is then similar to the quantum volume~\cite{Mol18} often used to characterize quantum hardware.

\section{Symmetry verification}
Our study is motivated by the presence of symmetries in quantum mechanical systems.
In such systems, one has a Hamiltonian $\hat{H}$, and is usually interested in studying the properties of ground or low-lying eigenstates of the system.
A (unitary) symmetry of a system is a unitary operator $\hat{S}$ that commutes with the Hamiltonian - $[\hat{H},\hat{S}]=0$.
When this is true, $\hat{H}$ may be block diagonalized within the eigenspaces of $\hat{S}$.
Then, if one were to study eigenstates of $\hat{H}$ on a quantum computer, one may perform such a study entirely within a single target eigenspace $\Ss$ of $\hat{S}$.
In real-world quantum computers, noise may shift the state of the computer outside of the target eigenspace $\Ss$.
By verifying during or at the end of a calculation that the system remains in $\Ss$, and throwing away results where this is not the case, it is thus possible to make our quantum computation less sensitive to these types of noise.

Verification of a symmetry is performed by measurement and post-selection which is typically performed in the computational basis (the eigenstates $|0\>$ and $|1\>$ of a single qubit).
The Pauli operators $\PP^N$ may be rotated into this basis relatively easily (see Sec.~\ref{sec:circuits}), and as such are a good class from which to draw symmetry operators.
If $\hat{S}\notin\PP^N$, but the target eigenspace $\Ss$ lies within the eigenspace of a Pauli operator $\hat{P}$, then measuring $\hat{P}$ presents a low-cost alternative to measuring $\hat{S}$, though this may provide less error mitigation in the case where the eigenspace of $\hat{P}$ is strictly larger than $\Ss$.
In general, symmetry verification will work with any construction of a projector valued measurement $\{\hat{M}_i\}$ where one projector $\hat{M}_S$ projects onto the target eigenspace $\Ss$.
We note that phase estimation~\cite{Kit96} provides a generic construction for such a measurement, although this is a rather high cost circuit (in particular requiring the ability to apply the symmetry $\hat{U}$ on the quantum computer).
This requirement for measurement implies that symmetry verification cannot be extended to antiunitary symmetries (nor to symmetries that anticommute with the Hamiltonian), as these do not lead to eigenspaces that can be projected into.

The projector valued measurement $\{\hat{M}_i\}$ is the more general object for symmetry verification than the symmetry $\hat{S}$.
In an arbitrary quantum circuit at an arbitrary time, if we know by any means that the state $|\psi\>$ in the absence of error satisfies $\hat{M}_s|\psi\>=|\psi\>$, measuring $\{\hat{M}_i\}$ on the noisy state $\rho$ and post-selecting will project to the state
\begin{equation}
\rho_s=\frac{\hat{M}_s\rho\hat{M}_s}{\Trace[\hat{M}_s\rho]}\label{eq:rhos_def}.
\end{equation}
Then, we have
\begin{equation}
\Trace[\rho_s|\psi\>\<\psi|]=\frac{\Trace[\rho|\psi\>\<\psi|]}{\Trace[\hat{M}_s\rho]}\geq \Trace[\rho|\psi\>\<\psi|]\label{eq:sym_ver},
\end{equation}
and our new state $\rho_s$ has strictly greater overlap with the target $|\psi\>$ than the pre-selection $\rho$ (unless $\hat{M}_s\rho\hat{M}_s=\rho$, in which case $\rho_s=\rho$).
Such a procedure can be immediately extended to multiple operators $\hat{S}_1,\hat{S}_2,\ldots$, as long as $[\hat{S}_i,\hat{S}_j]=0$. (If this is not the case, sequential symmetry verification projects between different eigenspaces, which is inefficient and greatly increases the number of experiments that must be thrown away.)
Symmetry verification may also be repeated at multiple points during a quantum circuit, by inserting measurement of $\hat{S}$ in between gates, as long as we expect the state $|\psi(t)\>$ to be an eigenstate of $\hat{S}$ at time $t$ during the circuit.
We call such protocols `bulk' symmetry verification, as opposed to `final' symmetry verification at the end of the an experiment.

\section{Ancilla and in-line symmetry verification}\label{sec:circuits}
\begin{figure}
\centering{
\includegraphics[width=\columnwidth]{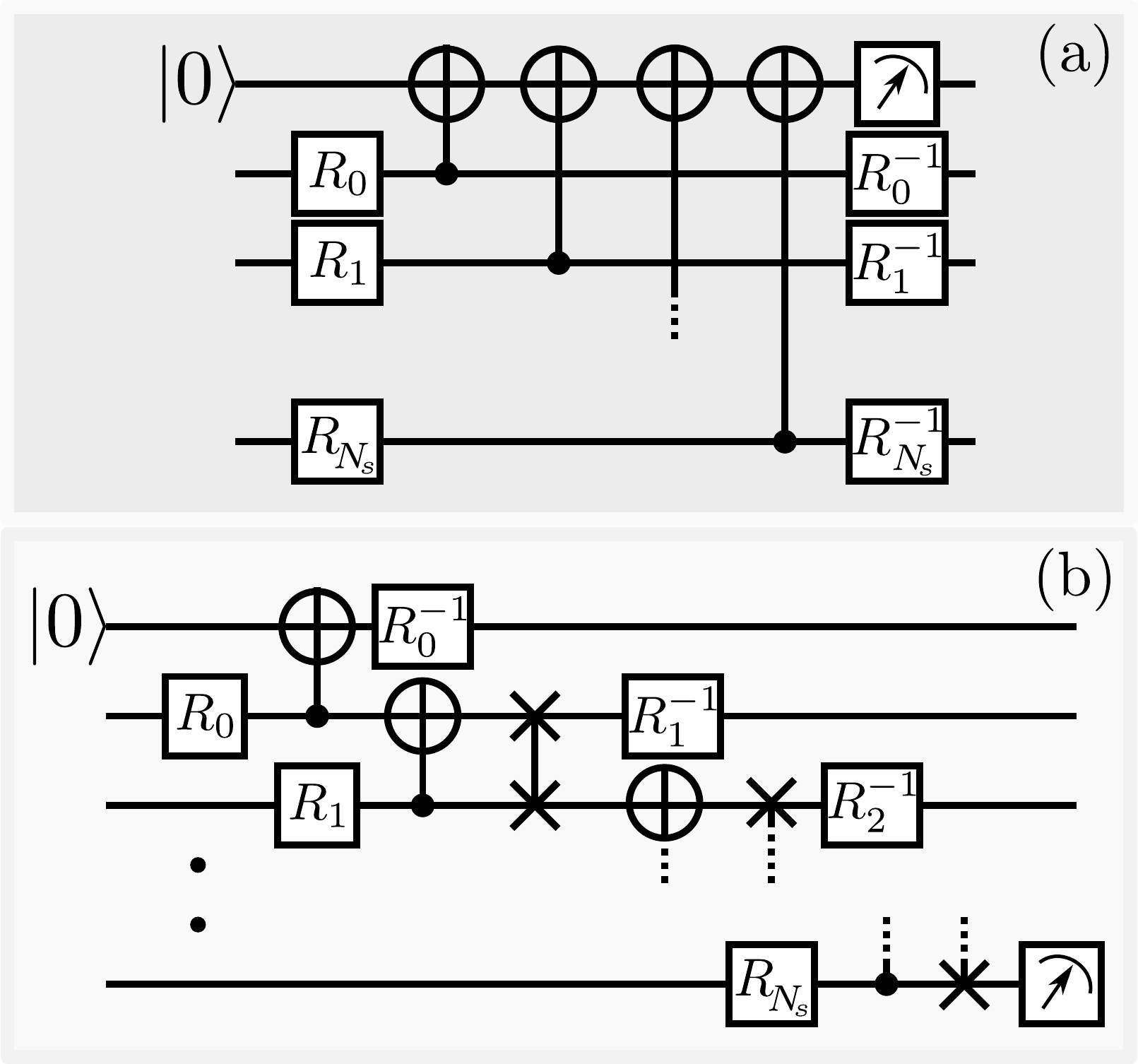}}
\caption{\label{fig:ancilla_simple} Quantum circuit for ancilla symmetry verification of a symmetry $\hat{S}$. (a) A simple circuit entangling all qubits with a single ancilla qubit. The rotations $\hat{R}_i$ depend on the tensor components $\hat{S}_i$ on each qubit $i$ (relationship given in text). (b) A circuit making an identical measurement to that in (a), but with only local CNOT and SWAP two-qubit gates. A SWAP between qubit 0 and the ancilla is not required because the Bell state prepared after the first CNOT is symmetric between the two qubits (this is not the case for the remaining qubits).}
\end{figure}

The simplest form of the symmetry verification involves the use of an ancilla qubit to measure the Pauli symmetry $\hat{S}$.
Let us write $\hat{S}\in\PP^n$ in terms of its tensor factors; $\hat{S}=\otimes_i\hat{S}_i$, and let $N_S$ be the number of nontrivial $\hat{S}_i=\{X,Y,Z\}$.
To each such $\hat{S}_i$, we can associate a corresponding rotation $\hat{R}_i=\{\exp(i\frac{\pi}{2}Y),\exp(-i\frac{\pi}{2}X),\mathbf{1}\}$ (such that $\hat{R}_i|\hat{S}_i=1\>=|0\>$).
The verification circuit is then shown in Fig.~\ref{fig:ancilla_simple}(a).
For each non-trivial $\hat{S}_i$, the corresponding qubit is rotated by $\hat{R}_i$, then performs a controlled-NOT gate on the ancilla qubit, and finally is rotated by $\hat{R}_i^{-1}$.
This requires that the ancilla qubit be coupled to each qubit in the system register that it measures, which is in general not possible in a quantum circuit.
As a low-cost alternative (Fig.~\ref{fig:ancilla_simple}(b)), the ancilla qubit may be shuffled along the system register via SWAP gates as it performs the controlled phase gate.
In either case, as the ancilla qubit must interact with each register qubit individually, the circuit depth must be $O(N_S)$

\begin{figure}
\includegraphics[width=\columnwidth]{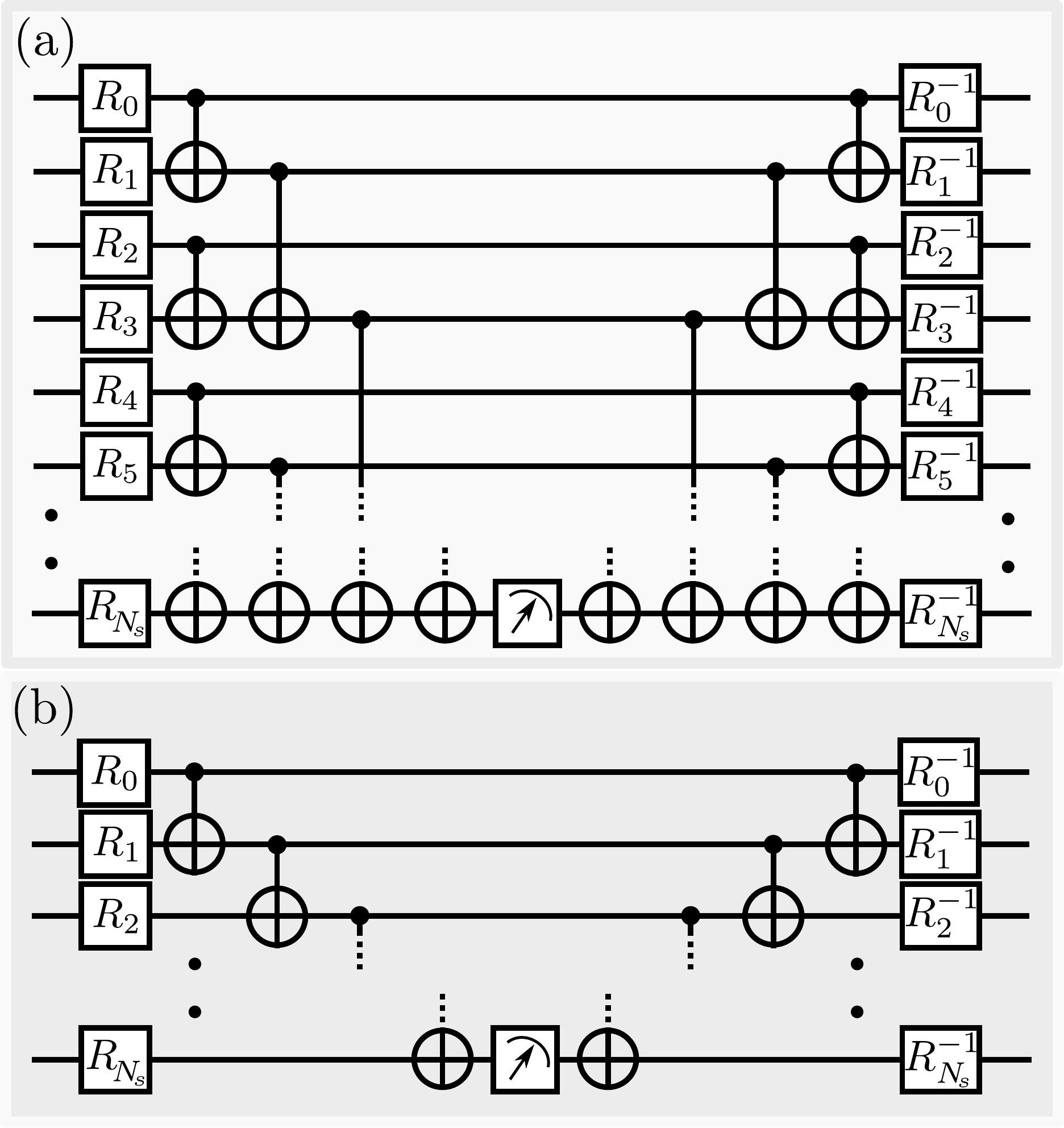}
\caption{\label{fig:in-line_verification} Quantum circuits for in-line symmetry verification. (a) The optimal verification circuit has $O(\log(N_{S}))$ depth, but requires long-range connectivity between qubits, which is not available on many architectures. (b) In the presence of linear connectivity, an $O(N_{S})$ depth verification circuit is optimal.}
\end{figure}

It is possible to forego the ancilla qubit in symmetry verification, by instead encrypting the symmetry $\hat{S}$ onto the computational degree of freedom of a qubit within the system itself, which is then read out.
In Fig.~\ref{fig:in-line_verification}(a) we give an example circuit for this in-line symmetry verification, with circuit depth only $O(\log(N_S))$.
This logarithmic depth requires qubits to be coupled as a binary tree, which is not possible in systems which allow only local couplings.
In general, for such a $d$-dimensional local coupling, the depth of the circuit must be at least $O(N_S^{1/d})$, being the minimum depth of a light-cone encompassing $N_S$ qubits.
In Fig.~\ref{fig:in-line_verification}(b) we give such a circuit for a system with linear connectivity.
Even when all-to-all coupling is available, the $O(\log(N_S))$-depth circuit (Fig.~\ref{fig:in-line_verification}(a)) may not be preferable, as the duty cycle for each qubit (i.e. the period of time between the first and last gate each qubit is involved in) is length $O(\log(N_S))$.
By contrast, the duty cycle of an individual qubit during the circuit in Fig.~\ref{fig:ancilla_simple}(b) is $O(1)$.
A short duty cycle implies that qubits can be used to perform other operations while the symmetry verification is ongoing, reducing the time cost when this circuit is performed as a small block of a larger computation.

\section{Variational quantum eigensolvers}
As an example target algorithm for symmetry verification, we consider ground state preparation for a Hamiltonian $\hat{H}$ via a variational quantum eigensolver~\cite{Per14,Mcc16}.
An (ideal) variational quantum eigensolver consists of a unitary circuit $U(\vec{\theta})$, parametrized by a vector of free angles $\vec{\theta}$ that control individual gates within the circuit.
This circuit acts on a starting state, which we take to be the computational basis state $|0,\ldots,0\>$, to produce a variational final state $|\psi(\vec{\theta})\>=U(\vec{\theta})|0,\ldots, 0\>$.
These angles are controlled classically to minimize the energy $E(\vec{\theta})=\<\psi(\vec{\theta})|\hat{H}|\psi(\vec{\theta})\>$.
This expectation value is calculated in an experiment by taking the Pauli decomposition of $\hat{H}$ (Eq.~\ref{eq:Pauli_decomp}), preparing $|\psi(\vec{\theta})\>$ and measuring each $\hat{P}_i$ repeatedly to accumulate statistics on $\<\psi|\hat{P}_i|\psi\>$.

Variational quantum eigensolvers (VQE) are natural candidates for final symmetry verification, and common classes of VQEs are also natural candidates for bulk symmetry verification.
In particular, for fermionic systems (such as the electronic structure problem), global fermion parity is conserved, making it a prime target for symmetry verification.
(At low energy, for non-superconducting systems, the particle number is often conserved as well, but this is not a Pauli operator, and is much more difficult to measure.)
Using the Jordan-Wigner transformation on an $N$-fermion Hamiltonian, this symmetry takes the form $Z^{\otimes N}$.
Most VQEs consist of creating an approximate starting state (such as the Hartree-Fock state) that respects this symmetry, and then performing multiple local rotations that continue to respect this symmetry.
This is true of both the unitary coupled cluster (UCC) ansatz~\cite{Per14}, and the quantum approximate optimization algorithm (QAOA)~\cite{Far14}.
In the former, the ansatz is taken as the expansion of the cluster operator $e^{\hat{T}-\hat{T}^\dag}$
\begin{align}
\hat{T}&=\sum_n\hat{T}^{(n)},\\
\hat{T}^{(n)}&=\sum_{i_1,\ldots,i_n;j_1,\ldots,j_n}\theta^{i_1,\ldots,i_n}_{j_1,\ldots,j_n}(\prod_{m=1}^n\chatdag_{i_m})(\prod_{m=1}^n\chat_{j_m})
\end{align}
where the $\theta$ parameters are taken as the free parameters to be optimized, and the sum is a sum over empty molecular orbitals to the left of the semi-colon, and filled molecular orbitals to the right.
This exponentiation is typically performed by the Trotter-Suzuki expansion, leaving a series of unitaries
\begin{equation}
\prod_{i;j}e^{\theta^i_j(\chatdag_i\chat_j-\chatdag_j\chat_i)}\prod_{i,j;k,l}e^{\theta^{i,j}_{k,l}(\chatdag_i\chatdag_j\chat_k\chat_l-\chatdag_l\chatdag_k\chat_j\chat_i)}\ldots
\end{equation}
each of which respects fermion parity.
QAOA for the electronic structure problem consists of performing steps of time evolution alternating between the Hartree-Fock Hamiltonian and the electronic-structure Hamiltonian, both of which respect fermion parity.
Thus, for both ansatz, bulk symmetry verification could be performed between individual steps of the time evolution.

Although symmetry verification promises a final state with greater overlap with the ground state, it does not promise a necessarily lower energy.
Let us write the (un-normalized) symmetry-accepted state $\rho_s$, and the symmetry-rejected state $\rho_r$. If our measurement was perfect, we would have
\begin{equation}
\rho_s=\hat{M}_{s}\rho\hat{M}_{s},\hspace{0.5cm}\rho_r=(\II-\hat{M}_{s})\rho(\II-\hat{M}_{s}).
\end{equation}
Then, $\Trace[\hat{H}\rho]=\Trace[\hat{H}\rho_r]+\Trace[\hat{H}\rho_s]$.
Now, suppose the rejected state $\rho_r$ has lower energy than the accepted state $\rho_s$;
\begin{equation}
\frac{\Trace[\hat{H}\rho_r]}{\Trace[\rho_r]}<\frac{\Trace[\hat{H}\rho_s]}{\Trace[\rho_s]}\label{eq:bad_sv}.
\end{equation}
We can calculate
\begin{align*}
&\Trace[\hat{H}\rho]=\Trace[\rho_r]\frac{\Trace[\hat{H}\rho_r]}{\Trace[\rho_r]}+\Trace[\rho_s]\frac{\Trace[\hat{H}\rho_s]}{\Trace[\rho_s]}\\
&<(\Trace[\rho_r]+\Trace[\rho_s])\frac{\Trace[\hat{H}\rho_s]}{\Trace[\rho_s]}=\frac{\Trace[\hat{H}\rho_s]}{\Trace[\rho_s]},
\end{align*}
and our symmetry-verified state would be higher in energy than the initial state as well.
As the energy of $\rho_r$ lies strictly above the ground state, failure of symmetry verification must imply $\rho_s$ has sufficiently large overlap with high-energy states.
As such, we would suggest that such a failure implies the energy of $\rho$ itself is not to be trusted.

\section{Post-selected symmetry verification and S-QSE}\label{sec:S-QSE}
Conveniently, when a quantum computation requires calculating the expectation values of a set of Pauli operators, symmetry verification may be performed via post-processing of the expectation values themselves (with possibly some additional measurements), rather than requiring additional quantum circuitry.
Suppose we want to calculate the expectation value of $\hat{P}\in\PP^N$ on our state $\rho$ following projection onto the $\hat{S}=s$($=\pm 1$) subspace of our symmetry $\hat{S}\in\PP^N$.
The projector onto this subspace may be written $\hat{M}_s=\frac{1}{2}(1+s\hat{S})$.
Then, the expectation value of $\hat{P}$ on the state $\rho_s$ targeted by the symmetry verification can be expanded using Eq.~\ref{eq:rhos_def}
\begin{align}
\Trace[\hat{P}\rho_s] &= \Trace \left[\hat{P} \frac{\hat{M}_s \rho \hat{M}_s}{\Trace[\hat{M}_s \rho]} \right] \nonumber\\
&= \frac{\Trace[\hat{P}\rho] + s\Trace[\hat{P}\hat{S}\rho]}{1 + s\Trace[\hat{S}\rho]},\label{eq:postselection}
\end{align}
where we have used the cyclic property of the trace and the fact that $[\hat{P},\hat{M}_s]=0$ to write $\Trace[\hat{P}\hat{M}_s\rho\hat{M}_s]=\Trace[\hat{P}\hat{M}_s\rho]$, and expanded our definition of $\hat{M}_s$.
The expectation values $\Trace[\hat{S}\rho]$, $\Trace[\hat{P}\rho]$, and $\Trace[\hat{P}\hat{S}\rho]$ may be then calculated using the unverified state $\rho$, and substituted into Eq.~\ref{eq:postselection} to obtain the verified result.
By avoiding additional quantum circuitry, we expect this method to outperform both ancilla and in-line symmetry verification.
However, we note that post-selection cannot be used for bulk symmetry verification (as we cannot measure these expectation values during the circuit).
Furthermore, it cannot be used in algorithms where the output is not an expectation value $\Trace[\hat{P}\rho]$.

Post-selected symmetry verification can be observed to be identical to a form of the quantum subspace expansion (QSE)~\cite{Mcc17}.
Originally designed to investigate the low-energy excited states around the ground space found by a variational quantum eigensolver, QSE works by taking a set of excitation operators $\{\hat{E}_i\}$, which can be applied to the approximated ground state $|\psi(\vec{\theta})\>$ to obtain a set of states $|\phi_i\>=\hat{E}_i|\psi(\vec{\theta})\>$.
The spectrum of the Hamiltonian within the manifold spanned by these states can be calculated as the solution to the generalized eigenvalue problem
\begin{equation}
\hat{H}_{\mathrm{QSE}}|\xi\>=\lambda\hat{B}_{\mathrm{QSE}}|\xi\>.
\end{equation}
Here, $\hat{H}_{\mathrm{QSE}}$ is the Hamiltonian matrix projected into the spanned manifold
\begin{equation}
[\hat{H}_{\mathrm{QSE}}]_{i,j}=\Trace[\hat{H}|\phi_i\>\<\phi_j|]\label{eq:HQSE_def},
\end{equation}
and $\hat{B}_{\mathrm{QSE}}$ is the overlap matrix,
\begin{equation}
[\hat{B}_{\mathrm{QSE}}]_{i,j}=\Trace[|\phi_i\>\<\phi_j|]\label{eq:BQSE_def},
\end{equation}
to account for the fact that $|\phi_i\>$ and $|\phi_j\>$ are in general not orthogonal.
In the presence of noise, although the state $|\phi_i\>$ is not well defined (as our noisy state $\rho$ is not a pure state), the operators $|\phi_i\>\<\phi_j|=\hat{E}_i\rho\hat{E}_j^{\dag}$ remain well-defined, and the expectation values in Eqs.~\ref{eq:HQSE_def} and~\ref{eq:BQSE_def} are still able to be measured in an experiment.

The set $\{\hat{E}_i\}$ is usually taken to be the set of low-order polynomials in qubit or fermion operators~\cite{Mcc17,Col18}.
However, if the set $\{\II,\hat{S}\}$ is chosen as excitation operators, the solution to the generalized eigenvalue problem is the same as that obtained by post-selection.
To show this, we expand
\begin{align}
\Trace[\hat{H}\rho_s]&=\sum_ih_i\Trace[\hat{P}_i\rho_s]\nonumber\\
&=\sum_i\frac{\Trace[h_i\hat{P_i}\rho] + s\Trace[h_i\hat{P_i}\hat{S}\rho]}{1 + s\Trace[\hat{S}\rho]}\nonumber\\
&=\frac{\Trace[\hat{H}\rho]+s\Trace[\hat{H}\hat{S}\rho]}{1+s\Trace[\hat{S}\rho]}\label{eq:rhos_trace}.
\end{align}
Next, we calculate the QSE matrices (using the commutation of $\hat{H}$ and $\hat{S}$)
\begin{align}
\hat{H}_{\mathrm{QSE}}&=\left[\begin{array}{cc} 
\Trace[\hat{H}\rho] & \Trace[\hat{H}\hat{S}\rho]\\
\Trace[\hat{H}\hat{S}\rho] & \Trace[\hat{H}\rho]
\end{array}\right],\\
\hat{B}_{\mathrm{QSE}} &=\left[\begin{array}{cc}
1 & \Trace[\hat{S}\rho]\\
\Trace[\hat{S}\rho] & 1
\end{array}\right]\label{eq:BQSE_calc}.
\end{align}
Assuming that $\Trace[\hat{S}\rho]\neq 1$, $\hat{B}_{\mathrm{QSE}}$ is invertible, the problem reduces to finding the (regular) eigenvalues of
\begin{equation}
\hat{B}_{\mathrm{QSE}}^{-1}\hat{H}_{\mathrm{QSE}}=\frac{1}{1-\Trace[\hat{S}\rho]^2}\left[\begin{array}{cc}
\alpha & \beta\\ \beta & \alpha
\end{array}\right],
\end{equation}
where
\begin{align}
\alpha &= \Trace[\hat{H}\rho]-\Trace[\hat{S}\rho]\Trace[\hat{H}\hat{S}\rho],\\
\beta &= \Trace[\hat{H}\hat{S}\rho]-\Trace[\hat{H}\rho]\Trace[\hat{S}\rho].
\end{align}
The eigenvalues of this matrix take the form
\begin{align}
\lambda&=\frac{1}{1-\Trace[\hat{S}\rho]^2}(\alpha\pm\beta)\\
&=\frac{\Trace[\hat{H}\rho]\pm\Trace[\hat{H}\hat{S}\rho]}{1\pm\Trace[\hat{S}\rho]},
\end{align}
which can be seen to be equal to those found in Eq.~\ref{eq:rhos_trace}.
We call this version of the quantum subspace expansion symmetry-QSE, or S-QSE for short.

This result is not surprising; it was suggested in~\cite{Mcc17} to account for symmetries during QSE by projecting $\hat{H}_{\mathrm{QSE}}$ and $\hat{B}_{\mathrm{QSE}}$ into the symmetry subspace, which achieves the same result as in the above.
However, this demonstrates that one may account for symmetries via a version of QSE without calculating the full linear response.
Moreover, this implies that S-QSE corrects for both coherent and incoherent errors that project out of the $\hat{S}=s$ subspace.
By contrast, QSE with an operator that anticommutes with the Hamiltonian can only correct coherent errors (see appendix).
S-QSE may be immediately combined with other forms of QSE, for example linear response QSE, by including both sets of operators as excitations.

\section{Simulation of symmetry verification on the hydrogen molecule}
To first investigate symmetry verification in a simple setting, we use a VQE to find the ground-state energy of H$_2$ on two qubits.
This follows previous experimental demonstrations~\cite{Per14,Oma16,Kan17,Col18}.
We take the STO-3G basis for H$_2$, which has four spin-orbitals, and convert this into a qubit Hamiltonian via the Bravyi-Kitaev transformation.
The four spin-orbitals require four qubits to represent them on, but in this representation the Hamiltonian is diagonal on two of the qubits, which may be removed.
The remaining two-qubit Hamiltonian takes the form
\begin{equation}
\hat{H}=h_{0}\II \II+h_{1}\II Z+h_{2} Z\II + h_{3}XX + h_{4}YY + h_{5}ZZ,\label{eq:H2Ham2q}
\end{equation}
where $h_i$ are sums of integrated two and four-body terms from the original electronic structure problem.
The calculation of these terms, and the Bravyi-Kitaev transformation itself, were performed using the psi4~\cite{psi4} and OpenFermion~\cite{Openfermion} packages.
The Hamiltonian can be seen to commute with the symmetry $\hat{S}=ZZ$.
Our ground state wavefunction has non-trivial overlap with the Hartree-Fock wavefunction, which is in the $ZZ=-1$ subspace; this is then our target subspace. 
We follow the unitary coupled cluster ansatz of~\cite{Oma16}, which consists of exciting our system to the $|01\>$ state, and performing the unitary rotation
\begin{equation}
\hat{U}(\theta)=e^{-i\theta X_0Y_1}.
\end{equation}
This unitary rotation may be decomposed using standard methods~\cite{Whi09}.
As described previously, the VQE procedure consists of fixing $\theta$, repeatedly preparing $|\psi(\theta)\>$ and measuring collections of terms in the Pauli decomposition of $\hat{H}$ until a good estimate of the energy $E(\theta)$ is found.
This is then repeated at varying $\theta$ as demanded by a classical optimizer until a minimum $E(\theta)$ is found~\cite{Per14}.

We compare the performance of the three symmetry verification protocols described previously as a final symmetry verification step.
The ancilla symmetry verification is performed in the same manner as Fig.~\ref{fig:ancilla_simple}(a).
The in-line symmetry verification is performed in a manner similar to Fig.~\ref{fig:in-line_verification}(a), but as this is final symmetry verification, we have no need to undo the symmetry measurement.
Instead, to measure the expectation value of a Pauli operator $\Trace[\rho\hat{P}]$, we can propagate $\hat{P}$ through the symmetry verification circuit~\cite{Got98} and measure the corresponding Pauli term.
It is then sufficient to rotate the control qubit to recover the expectation values $\<\II Z\>$ and $\<XX\>$.
From this we may calculate all other expectation values in Eq.~\ref{eq:H2Ham2q} using the fact that $ZZ=-1$.
For this problem, S-QSE not only requires no additional circuitry, but also no additional measurements (all required terms are in the Pauli decomposition of the Hamiltonian).

To test symmetry verification in the presence of realistic noise, we simulate our chosen experiment using the quantumsim density matrix simulator~\cite{quantumsim}.
We take gate error models and parameters similar to previous simulation work based on experimental data of state-of-the-art superconducting transmon qubits~\cite{Obr17}.
Errors in transmon qubits are dominated by decoherence times, which we take at a base level to be $T_1=T_2=20~\mu$s.
This should be compared to single and two-qubit gate times of $20$~ns (giving a total circuit length without symmetry verification of $220$~ns).
Single and two-qubit gates suffer from additional dephasing noise of $0.01\%$ and $1\%$ respectively.
We assume that single-shot measurement (for verification purposes) has a read-out error of $1\%$, and that error in tomographic measurements and pre-rotations (used to calculate the expectation values themselves) can be cancelled by linear inversion tomography~\cite{Filipp09,Chow10}.

\begin{figure}
\includegraphics[width=\columnwidth]{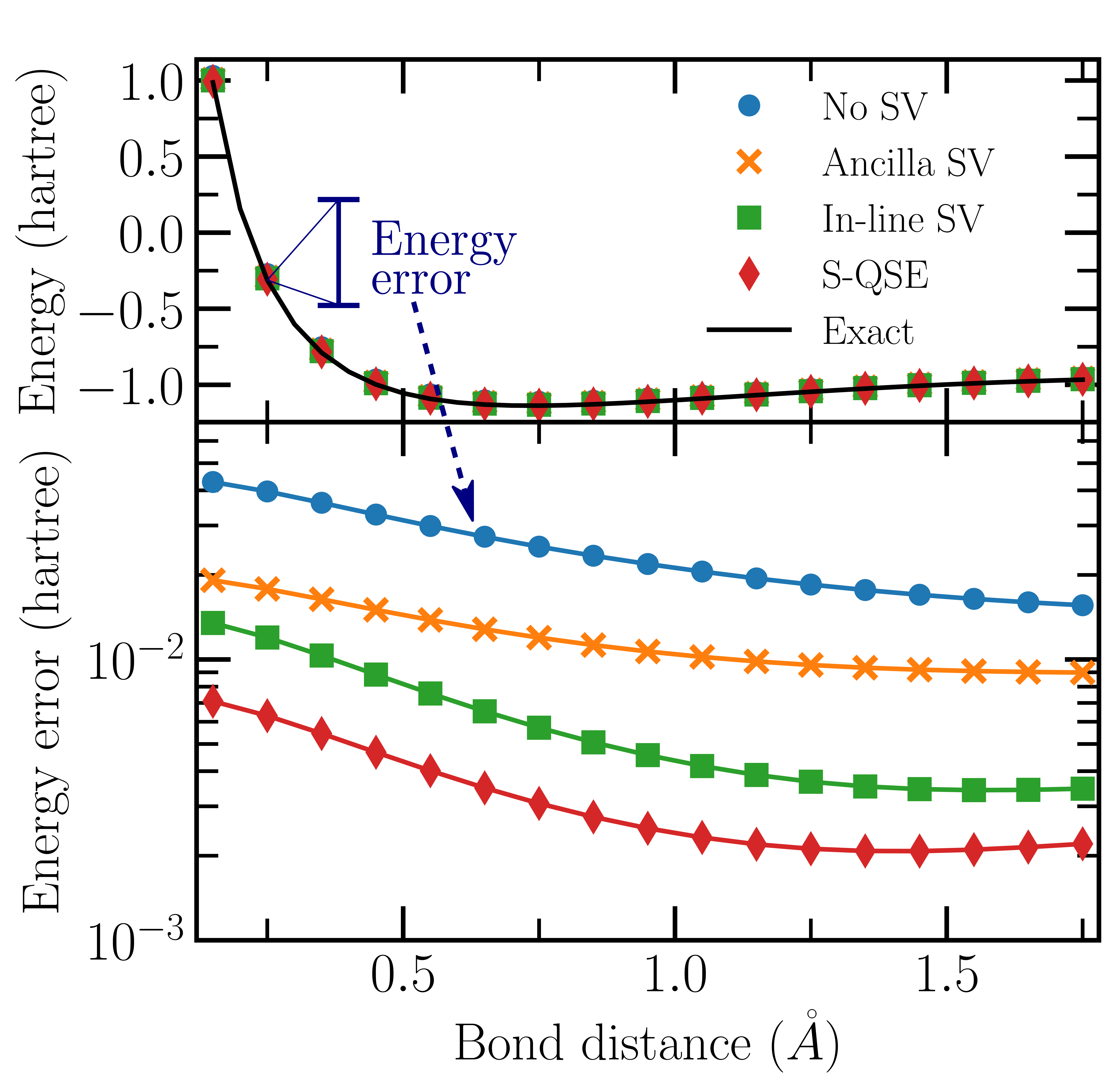}
\caption{\label{fig:comp}(Color online) Accuracy of the VQE over the entire bond dissociation curve using the different symmetry verification methods mentioned in the text (labelled in legend). (top) The target curve of H$_2$, compared to the exact result (black line). (bottom) Log plot of the difference between the black lines and points in the above plot.}
\end{figure}

Using the above error model, we observe (Fig.~\ref{fig:comp}) that the un-mitigated VQE (blue points) achieves an error in the energy of approximately $0.01-0.04$ hartree across the bond dissociation curve.
This error is improved upon by all symmetry verification techniques.
S-QSE (red diamonds) provides the largest improvement of all symmetry verification protocols, as no additional errors are introduced.
The S-QSE circuit is observed to give approximately a five-fold improvement over the unmitigated circuit, while ancilla (orange crosses) and in-line (green squares) symmetry verification show an approximately two-fold and three-fold improvement respectively.
The differences between S-QSE and other forms of symmetry verification emphasize the importance of minimizing the verification cost in bulk symmetry verification (where S-QSE is no longer available).

\begin{figure}
\includegraphics[width=\columnwidth]{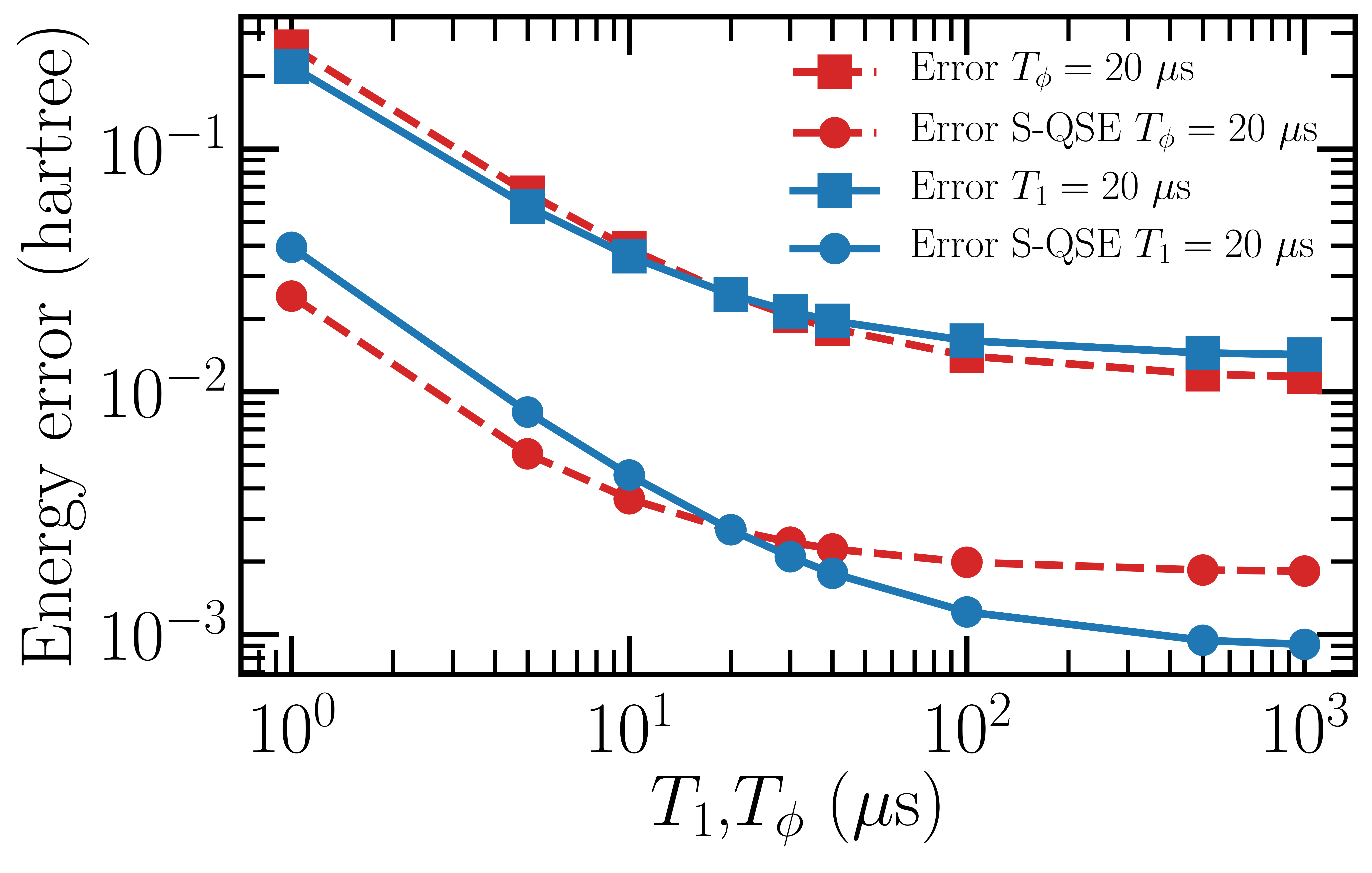}
\caption{\label{fig:err_rate}(Color online) Effect of varying decoherence times on the VQE accuracy. With all other error sources turned off, $T_1$ is varied with $T_{\phi}=20~\mu$s fixed (red-dashed curves), and $T_{\phi}$ is varied with $T_{1}=20~\mu$s fixed (blue-solid curves). We plot the error in estimating the ground-state energy for the unmitigated experiment (squares), and the circuit mitigated with S-QSE (circles). Data points for the blue and red curves are identical at $T_1=T_{\phi}=20~\mu$s, as can be seen from the complete overlap.}
\end{figure}

We now investigate the effect of different noise channels on the performance of symmetry verification.
Any noise channel that commutes with the symmetry operators evolves the system state within the target subspace, which symmetry verification explicitly does not mitigate.
The analysis of which channels have this property can be reduced to an analysis over $\PP^N$, as if we mitigate Pauli errors $\hat{P}_i\in\PP^N$, we also mitigate any linear combination of them~\cite{Got09}.
In the above circuit, the $ZZ$ symmetry commutes with any single-qubit $Z$ errors, making the protocol prone to the $T_{\phi}$ (pure dephasing) channel, but it anticommutes with single-qubit $X$-errors, making the protocol resilient against the $T_1$ (amplitude decay) channel.
To investigate this, in Fig.~\ref{fig:err_rate} we calculate the error in determining the ground state energy near the minima of the bond dissociation curve ($0.75$\r{A} bond distance) using S-QSE, as we vary $T_1$ and $T_{\phi}$.
We turn all other error sources off, and vary $T_1$ ($T_{\phi}$) with $T_{\phi}=20~\mu$s ($T_1=20~\mu$s) fixed.
In the absence of error mitigation, the two decoherence sources have almost identical effect (deviation approximately $10^{-2}$ hartree).
However, in the presence of error mitigation, the susceptibility of the VQE to $T_{1}$ noise is noticeably smaller than to $T_{\phi}$ noise - up to a factor of two over the range of decoherence times plotted.
We note that S-QSE does not make our circuit second-order sensitive to $T_1$ noise.
This can be understood as X-errors at some points during our VQE circuit are rotated to Z-errors by later gates in the circuit, preventing their mitigation.

\section{Inserting and rotating symmetries}\label{sec:insertion}

As observed in the previous section, verifying single symmetries has a marked effect on the performance of a quantum circuit, but will not catch and remove all sources of noise.
In this section we suggest how one may improve upon this by adding additional symmetries to the quantum algorithm, and by rotating existing symmetries to make them more sensitive to errors on the underlying quantum hardware.
In the language of quantum error correction, this is a low-cost attempt to increase the distance of the detection code.

We first suggest a method to extend an $N$-qubit Hamiltonian $\hat{H}$, given a Pauli operator $\hat{P}\in\PP^N$, to an $N+1$-qubit Hamiltonian $\hat{H}_{\mathrm{ext}}$
\begin{equation}
\hat{H}_{\mathrm{ext}}=\left[\begin{array}{cc}\hat{H} & 0 \\ 0 & \hat{P}\hat{H}\hat{P}\end{array}\right].
\end{equation}
Both blocks of $\hat{H}_{\mathrm{ext}}$ can be seen to have the same eigenspectrum (as this is unaffected by the unitary rotation of $\hat{P}$), and $\hat{H}_{\mathrm{ext}}$ commutes with the operator
\begin{equation}
\left[\begin{array}{cc}0&\hat{P}\\\hat{P}&0\end{array}\right]=X\hat{P},
\end{equation}
which is then the new symmetry operator.
This mapping corresponds to mapping Pauli operators $\hat{Q}\in\PP^N$ in the original problem to
\begin{equation}
\hat{Q}_{\mathrm{ext}}=\begin{cases}
\II\hat{Q}&\mathrm{if}\; [\hat{Q},\hat{P}]=0\\
X\hat{Q}&\mathrm{if}\; \{\hat{Q},\hat{P}\}=0\end{cases}\label{eq:pauli_addition}.
\end{equation}
To implement this in the algorithm itself, we note that every circuit can be decomposed into a product of unitary rotations
\begin{equation}
\prod_je^{i\theta_j\hat{Q}_j},\hspace{1cm}\hat{Q}_j\in\PP^N\label{eq:circuit_decomp},
\end{equation}
where a single $\hat{Q}\in\PP^N$ may be repeated in the product.
Adding the symmetry then consists of replacing these rotations by rotations around the transformed operator $\hat{Q}_{\mathrm{ext}}$ (as per Eq.~\ref{eq:pauli_addition}), and re-decomposing the operations into a circuit (using e.g.~the methods of~\cite{Whi09,Has15}).
If $\hat{H}$ had a previous set of symmetries $\hat{S}_i$, these are transformed to a new set $\hat{S}_{i,\mathrm{ext}}$ (following Eq.~\ref{eq:pauli_addition}), that commute with both $\hat{H}_{\mathrm{ext}}$ and the additional symmetry $X\hat{P}$.
This extension method is particularly suitable for digital quantum simulation, as circuits are often generated in the form of Eq.~\ref{eq:circuit_decomp}.
This is the case for traditional Hamiltonian simulation~\cite{Geo14}, quantum phase estimation~\cite{Whi09}, and the UCC QSE discussed previously, all of which require exponentiating an operator via the Suzuki-Trotter expansion~\cite{Suz76}.

Beyond choosing the number of symmetries in a problem, one may wish to choose how these symmetries appear in the problem.
In particular, sets of symmetries may be found that anticommute with all local operators, which should increase the mitigation power of the verification protocol against local sources of noise.
(For example, the $N$-qubit operators $X^{\otimes N}$ and $Z^{\otimes N}$ with even $N$.)
Any two groups of $M$ Pauli operators are unitarily equivalent as long as they satisfy the same commutation and multiplication rules (e.g. $\II Z$, $Z\II$, and $ZZ$ are equivalent to $XX$, $YY$ and $ZZ$, but not to $\II X$, $\II Y$ and $\II Z$).
To find such unitary transformations, we suggest decomposing them into rotations of the form $\hat{R}=e^{i\frac{\pi}{2}\hat{Q}}$ for $\hat{Q}\in\PP$, which transforms
\begin{equation}
\hat{P}\in\PP\rightarrow \hat{R}^{\dag}\hat{P}\hat{R}=\begin{cases}\hat{P}&\mathrm{if}\;[\hat{P},\hat{Q}]=0\\i\hat{P}\hat{Q}&\mathrm{if}\;\{\hat{P},\hat{Q}\}=0\end{cases}\;.
\end{equation}
Rotations of this form have a few desirable properties.
Their effect is easy to calculate classically, and they transform Pauli operators to Pauli operators.
Furthermore, each $\hat{R}$ leaves half of the Pauli group unchanged.
This allows for some choice of rotations to leave desired symmetries (or other operators) already present in the problem invariant, while other terms are rotated.

\section{Extending the symmetry verification of the hydrogen molecule}

We now demonstrate the verification of multiple symmetries by extending the previous VQE simulation of H$_2$.
We transform the electronic structure Hamiltonian onto a qubit representation this time via the Jordan-Wigner transformation.
This gives the four-qubit Hamiltonian
\begin{align}
\hat{H}=&h_I\II +\sum_ih_iZ_i +\frac{1}{2}\sum_{i\neq j}h_{i,j}Z_iZ_j\nonumber\\&+h_s(X_0Y_1Y_2X_3+Y_0X_1X_2Y_3\nonumber\\&\hspace{1cm}-X_0X_1Y_2Y_3-Y_0Y_1X_2X_3),
\end{align}
which has symmetries $\hat{S}_0=Z_0Z_1$, $\hat{S}_1=Z_0Z_2$, and $\hat{S}_2=Z_0Z_1Z_2Z_3$.
In the Bravyi-Kitaev transformation these symmetries were the additional qubits that were thrown away.
We choose again the unitary coupled cluster ansatz for the VQE, which can be reduced to the operator\footnote{The cluster operator for this system is a sum of 8 four-qubit terms, however the action of each term on the Hartree-Fock starting state is identical, so only one is needed.}
\begin{equation}
\hat{U}(\theta)=e^{i\theta Y_0X_1X_2X_3}.
\end{equation}
As in the two-qubit case, the VQE circuit consists of preparing the system in the Hartree-Fock state $|1100\>$, applying $U(\theta)$ and measuring the variational energy, for a total circuit time of $400$~ns.

The above set of symmetries still commute with all single-qubit $Z$ errors, so we rotate our problem to increase the mitigation power of symmetry verification.
We choose the rotation
\begin{equation}
\hat{R}=e^{i\frac{\pi}{2}Y_0X_2}e^{i\frac{\pi}{2}Y_1X_3}.
\end{equation}
This transforms the symmetry $\hat{S}_0\rightarrow X_0X_1X_2X_3$, whilst leaving $\hat{S}_1$ and $\hat{S}_2$ unchanged.
The resulting set of symmetries do not commute with any single-qubit $X$ or $Z$ operator, as required.
To create the transformed circuit, we need to transform both our starting state $|1100\>\rightarrow\hat{R}|1100\>$, and the UCC unitary ansatz 
\begin{equation}
\hat{U}(\theta)\rightarrow\hat{R}\hat{U}\hat{R}^{\dag}=e^{i\theta Y_0Z_1X_2}.
\end{equation}
The transformed circuit incurs an additional cost from this initial application of $\hat{R}$, but this is balanced by the reduced weight of the transformed cluster operator, resulting in a total circuit time of $440$~ns.

\begin{figure}
\includegraphics[width=\columnwidth]{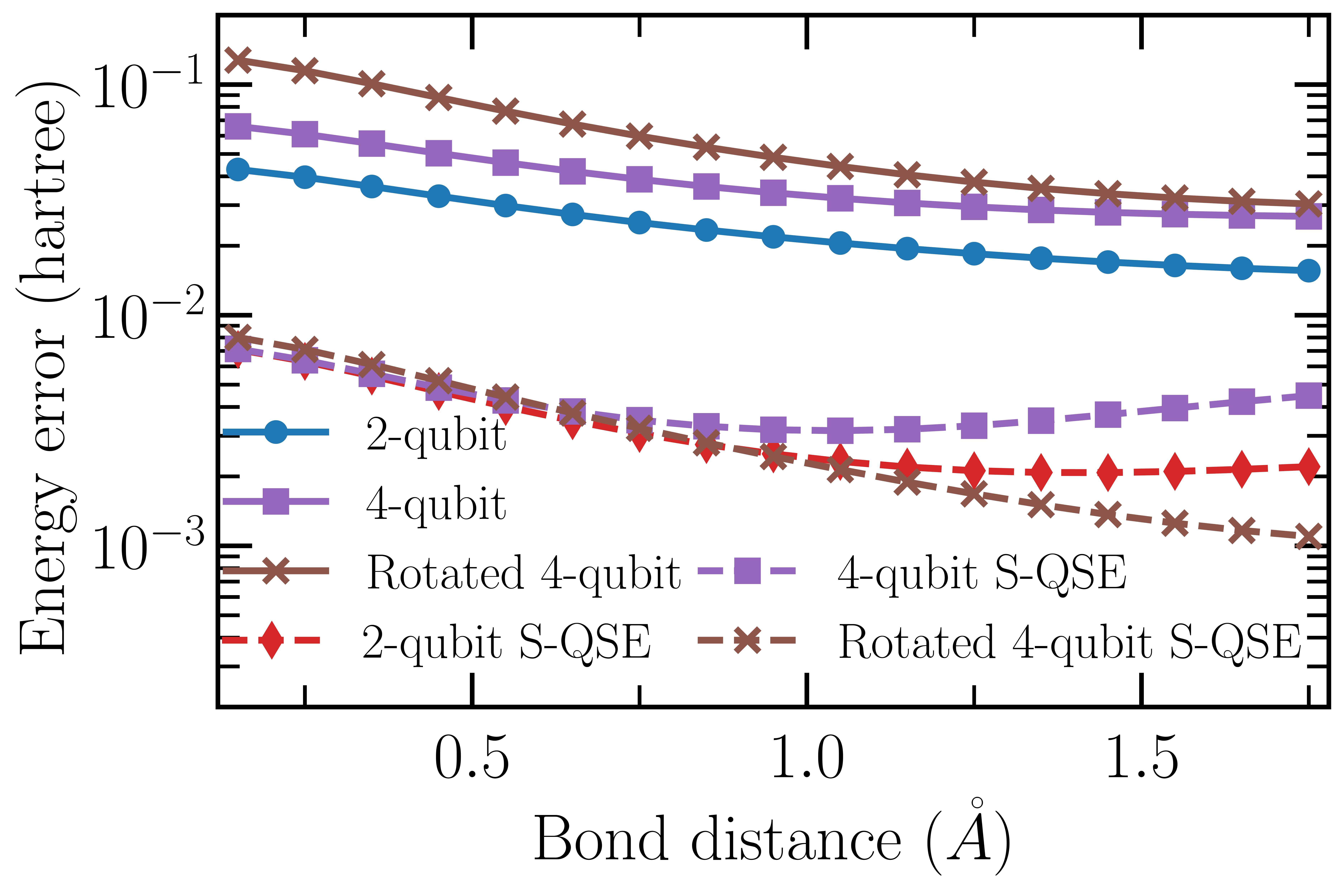}
\caption{\label{fig:4q_comp}(Color online) Adding and adjusting symmetries to optimize symmetry verification. The blue (dots) and red (diamonds) curves correspond to their coloured (shaped) counterparts in Fig.~\ref{fig:comp}, whilst the purple (squares) and brown (crosses) curves come from a four-qubit simulation of H$_2$ using the two protocols described in the text. 
The dashed lines represent the S-QSE versions of their solid counterpars.
Error parameters on all qubits are the same for all simulations (parameters given in the text).}
\end{figure}

In Fig.~\ref{fig:4q_comp}, we compare the performance of the two different circuits above to the two-qubit circuit of Fig.~\ref{fig:comp}, with and without the addition of S-QSE.
At small bond distance ($\lesssim 0.75$\r{A}), the target ground state (in the absence of rotation by $\hat{R}$) is roughly a computational basis state, which is immune to dephasing errors.
At this point, all three verification protocols perform roughly similarly, despite the unmitigated four-qubit simulations performing significantly worse than the unmitigated two-qubit simulation.
At large bond distance ($\gtrsim 0.75$\r{A}), the ground state is prone to $T_2$ noise, at which point we see the rotated 4-qubit S-QSE simulation significantly outperforming its counterparts.
At the largest distance studied, this simulation achieves a two-fold reduction in error compared to the two-qubit S-QSE simulation, despite using twice as many qubits and a twice as long circuit.
By comparison, unrotated S-QSE on four qubits cannot protect against the $T_2$ noise accumulated over the simulation, and performs a factor of two worse than the two-qubit S-QSE simulation.
This clearly demonstrates the need to optimize symmetry verification protocols to account for errors present in the system as this technique is scaled up to larger computations.
Over the entire bond-dissociation curve, the rotated four-qubit S-QSE simulation outperforms its unmitigated counterpart by over an order of magnitude.

\section{Conclusion}

In this paper we have presented a new low-cost strategy for error mitigation, which we call symmetry verification. We have discussed various ways in which it can be applied to different algorithms, and various methods to optimize the mitigation power against common sources of error. We have demonstrated these protocols on a simulated VQE experiment of H$_2$, and observed that they outperform the unmitigated result over the entire bond-dissociation curve by around an order of magnitude.

Although the above techniques are very promising for small experiments, much work needs to be done optimizing symmetry verification for mid-range experiments in the NISQ era.
The addition and choice of symmetries needs to be investigated further to minimize the resulting circuit depth.
Further study is also needed on the optimal number of symmetry verifications to be added to a circuit, both to maximise mitigation and minimize run-time (which increases exponentially in the number of verifications made).
Finally, given the obvious connection between symmetry verification and the stabilizer formalism of quantum error correction, it is natural to ask whether one can mix the two to transform slowly between mid-size NISQ circuits and large-scale fault-tolerant ones.

While this paper was in production, a related work by McArdle et al.~\cite{Mca18} appeared on the ArXiv.
They simulate the performance of ancilla symmetry verification for a VQE, and its combination with other error mitigation strategies to further improve robustness against noise.
Their results are consistent with and complementary to our own, and they provide useful techniques for measuring non-Pauli operators not considered in this work.

\acknowledgments{The authors wish to thank Carlo Beenakker, Leonardo DiCarlo, Brian Tarasinski, Barbara Terhal, Detlef Hohl, Luuk Visscher, Francesco Buda, Yaroslav Herasymenko, and Adriaan Rol for feedback, advice and support in this project. This research was supported by the Netherlands Organization for Scientific Research (NWO/OCW) and by an ERC Synergy Grant.}

\appendix
\section*{Appendix: Error mitigation of QSE with anticommuting operators}\label{app:QSEanticommuting}
In this appendix we repeat the analysis of QSE from the text, but with an operator $\hat{A}$ that anti-commutes with the Hamiltonian $\hat{H}$.
Let us assume to begin that $\hat{A}$ is unitary.
Such an operator cannot be simultaneously diagonalized with $\hat{H}$, and so we have no result from symmetry verification to compare with.
Given an eigenstate $\hat{H}|\psi\>=E|\psi\>$, we have that $\hat{H}\hat{A}|\psi\>=-\hat{A}\hat{H}|\psi\>=-E\hat{A}|\psi\>$, and so the presence of an anticommuting operator splits the eigenstates of $\hat{H}$ into pairs of equal magnitude but opposite sign energies (known as eigenstates of different chirality).
If $\hat{A}=\hat{A}^{\dag}$, the eigenstates of $\hat{A}$ itself are the equal superpositions 
\begin{equation}
|\pm\>=\frac{1}{\sqrt{2}}(|\psi\>\pm\hat{A}|\psi\>)\label{eq:pm_def}.
\end{equation}

For QSE, we must calculate the operators $\hat{H}_{\mathrm{QSE}}$ and $\hat{B}_{\mathrm{QSE}}$.
\begin{align}
\hat{B}_{\mathrm{QSE}} &=\left[\begin{array}{cc}
1 & \Trace[\hat{A}\rho]\\
\Trace[\hat{A}^{\dag}\rho] & 1
\end{array}\right].\\
\hat{H}_{\mathrm{QSE}}&=\left[\begin{array}{cc} 
\Trace[\hat{H}\rho] & \Trace[\hat{H}\hat{A}\rho]\\
\Trace[-\hat{H}\hat{A}^{\dag}\rho] & -\Trace[\hat{H}\rho]
\end{array}\right].
\end{align}
Again assuming $|\Trace[\hat{A}\rho]|^2\neq 1$, we can invert $\hat{B}_{\mathrm{QSE}}$ and calculate
\begin{equation}
\hat{B}_{\mathrm{QSE}}^{-1}\hat{H}_{\mathrm{QSE}}=\frac{1}{1-|\Trace[\hat{A}\rho]|^2}\left[\begin{array}{cc}
\alpha & \beta\\ -\beta^* & -\alpha^*
\end{array}\right],
\end{equation}
where
\begin{align}
\alpha &= \Trace[\hat{H}\rho]+\Trace[\hat{A}\rho]\Trace[\hat{H}\hat{A}\rho]\\
\beta &= \Trace[\hat{H}\hat{A}\rho]+\Trace[\hat{H}\rho]\Trace[\hat{A}\rho].
\end{align}
The solution to the equation is
\begin{align}
E_{\mathrm{QSE}}^2&=\frac{|\alpha|^2+|\beta|^2}{(1-|\Trace[\hat{A}\rho]|^2)^2}\\
&=\frac{\Trace[\hat{H}\rho]^2+|\Trace[\hat{H}\hat{A}\rho]|^2}{1-|\Trace[\hat{A}\rho]|^2}.
\end{align}
To understand the gain in energy, $|\Trace[\hat{H}\hat{A}\rho]|^2$, let us first consider a single set of opposite chirality states $|\psi\>$ and $\hat{A}|\psi\>$ (with energy $\pm E$).
We first note that if $\rho$ is an incoherent superposition of the eigenstates,
\begin{equation}
\rho=|a|^2|\psi\>\<\psi|+|b|^2\hat{A}|\psi\>\<\psi|\hat{A},
\end{equation}
$\Trace[\hat{H}\hat{A}\rho]=\Trace[\hat{A}\rho]=0$ (as $\<\psi|A|\psi\>=0$), and QSE strictly does not improve on the estimate of the ground state energy.
We next consider the opposite situation, where $\rho$ is a coherent superposition of eigenstates:
\begin{align}
\rho=&(\cos(\theta)|\psi\>+\sin(\theta)e^{i\phi}\hat{A}|\psi\>)\nonumber\\&\hspace{1cm}\times(\cos(\theta)\<\psi|+\sin(\theta)e^{-i\phi}\<\psi|\hat{A}^{\dag}).
\end{align}
We can calculate
\begin{align}
&\Trace[\hat{H}\rho]=E\cos(2\theta),\\
&\Trace[\hat{A}\rho]=\sin(2\theta)(1+Ae^{i\phi}),\\
&\Trace[\hat{H}\hat{A}\rho]=E\sin(2\theta)(Ae^{i\phi}-1),
\end{align}
where $A=\<\psi|\hat{A}^2|\psi\>$ (so $|A|\leq 1$, and for $\hat{A}\in\PP^N$, $A=1$).
This gives
\begin{align}
E_{\mathrm{QSE}}^2&=E^2\frac{\cos^2(2\theta)+\sin^2(2\theta)\chi_+}{1-\sin^2(2\theta)\chi_-},\\
\chi_{\pm}&=(1\pm Ae^{i\phi})(1\pm Ae^{-i\phi}).
\end{align}
We see that if $A=1,\phi=\frac{\pi}{2}$, QSE corrects the coherent error entirely, whilst if $A=1,\phi=0$ it has no effect.
This implies that QSE cannot correct coherent rotations of $\rho$ from $|\psi\>$ towards an eigenstate of $\hat{A}$.
This is in keeping with the general observations in~\cite{Mcc17} for the performance of QSE as an error mitigation strategy.

If $\hat{A}$ is not unitary, then $\hat{A}^{\dag}\hat{A}$ is a Hermitian operator that commutes with $\hat{H}$.
Importantly, if $\{\hat{A},\hat{H}\}=0$, $\{\hat{A}\hat{H},\hat{H}\}=0$ as well, giving a second anticommuting operator that is in general non-unitary.
This could be used directly in QSE, although the analysis of Sec.~\ref{sec:S-QSE} no longer holds unless $\hat{A}^{\dag}\hat{A}\in\PP^2$.
For symmetry verification, we require the form of the projector $\hat{M}_{a}$ onto the correct $\hat{A}^{\dag}\hat{A}|\psi\>=a|\psi\>$ subspace.
This is a difficult task in general to construct (for $\hat{A}\hat{H}$, it is equivalent to diagonalizing the Hamiltonian).
We have been unable to construct any further bounds on the performance of QSE as an error mitigation strategy for a general Hermitian operator, nor for an operator which neither commutes nor anti-commutes with $\hat{H}$. This is, however, an interesting direction for future research.

\bibliographystyle{apsrev4-1}
\bibliography{paper_resub.bbl}

\end{document}